\documentclass[aps,preprint,showpacs,superscriptaddress,groupedaddress]{revtex4-1}

\usepackage{graphicx}  
\usepackage{dcolumn}   
\usepackage{bm}        
\usepackage{amssymb}   
\usepackage{amsmath}
\usepackage{epsfig}
\usepackage[T1]{fontenc}
\usepackage{appendix}
\usepackage{color}
\usepackage{cleveref}
\usepackage[utf8]{inputenc}

\begin{document}

\title{Non-linear Synthesis of Complex Laser Waveforms at Remote Distances}

\author{Nicolas Berti$^{1}$}
\author{Wahb Ettoumi$^{1}$}
\author{Sylvain Hermelin$^{1}$}
\author{J\'{e}r\^{o}me Kasparian$^{2}$}\email{j\'{e}r\^{o}me.kasparian@unige.ch}
\author{Jean-Pierre Wolf$^{1}$}

\affiliation{$^{1}$Universit\'e de Gen\`eve, GAP-Biophotonics, Chemin de Pinchat 22, 1211 Geneva 4, Switzerland}
\affiliation{$^{2}$Universit\'e de Gen\`eve, GAP-Nonlinear, Chemin de Pinchat 22, 1211 Geneva 4, Switzerland}

\date{\today}


\begin{abstract}
{Strong deformation of ultrashort laser pulse shapes is unavoidable when delivering high intensities at remote distances due to non-linear effects taking place while propagating. 
Relying on the reversibility of laser filamentation, we propose to explicitly design laser pulse shapes so that propagation serves as a non-linear field synthesizer at a remote target location. Such an approach allows, for instance, coherent control of molecules at a remote distance, in the context of standoff detection of pathogens or explosives.}
\end{abstract}


\maketitle



\section{Introduction}
Manipulating atoms and molecules using quantum control or pump-probe excitation schemes requires precisely tailored electromagnetic fields~\cite{Scully2002,Rabitz2000,Krause1993,Wipfler2012,Dudovich2002} which require stability in time and space.
Most of the manipulations rely on non-linear interactions of broadband pulses with molecules, and therefore  high peak intensity (GW$\,$cm$^{-2}$ to tens of TW$\,$cm$^{-2}$) and ultrashort (femtosecond) pulses.
Therefore, the pulse propagation from the laser source to a remote target location is likely to encounter non-linear spatio-temporal perturbations as, e.g., Kerr-induced wavefront distortion, self-phase modulation and temporal pulse shape alteration, in addition to linear perturbations such as diffraction, dispersion and aberrations that stem from air turbulence. In laboratory-scale experiments, these perturbations can be handled using deformable mirrors and temporal or spectral pulse shapers. Linear deformations can be corrected over large distances as well. This is routinely performed with adaptive wavefront corrections based on artificial guide stars for large astronomical telescopes~\cite{HubinN1993}, or the correction of turbulence perturbations with temporal pulse shapers \cite{ExterBBBKW2008}. These corrections can be either static or dynamic, e.g., based on a feedback loop~\cite{Zhu2007}.  

In contrast, non-linear spatio-temporal phase shifts accumulate and give rise to dramatic modifications of the laser waveforms. If Kerr self-focusing overcomes diffraction, which requires the peak power to exceed a critical power of 3~GW, the beam can even experience a catastrophic collapse. The arrest of the collapse by self-defocusing processes (ionization or higher-order optical non-linearities~\cite{VincoB2004,ShimSHVHIG2010,Bejot}) leads to filamentation \cite{BraunKLDSM1995,Chin2005,Kasparian,CouaironM07}. The complex propagation dynamics associated to this highly non-linear propagation mode results in a dramatic reshaping of the pulse, challenging any explicit design of the initial pulse required for open loop approaches. Furthermore the lack of feedback from a remote pulse characterization at the target location prevents a closed-loop approach.

Active remote sensing, including standoff detection of explosives or harmful molecular agents, greatly benefits from non-linear interactions induced by ultrashort high intensity lasers. In particular, coherent Raman spectroscopy \cite{NatanLGKS2012,Bremer2011,Wipfler2012,Bremer2013,Glenn2014,Hemmer2011,Ariunbold2014,Motzkus2007}, multi-photon excited fluorescence \cite{M'ejKYFSW2004,Gravel2004,Boutou_FMN2}, and even atmospheric lasing \cite{ScullyPNAS_2011,ScullyPNAS_2012,Dogariu2011,Kartashov2013} require a precise and stable spatio-temporal waveform at a defined distance. A particularly perturbation-sensitive strategy is the discrimination of almost identical molecules using coherently controlled multiphoton fluorescence \cite{Assion_science1998,Boutou_FMN2,Boutou_FMN,Clow2009,Petersen2010,Tseng2011}, like optimal dynamic discrimination (ODD). In order to transpose the successful demonstration of  bacteria or biomolecules discrimination from the laboratory to an atmospheric standoff detection workframe, non-linear phase perturbations have to be included when manipulating optical waveforms: If the B-integral ($B \equiv k_{0} \int_{0}^{z} n_{2} I \textrm{d}z$ \cite{CouaironM07}) characterizing the accumulated non-linear phase shift exceeds a fraction of $\pi$, these perturbations have to be pre-compensated. Since the required laser power often exceeds the critical power in order to propagate over large distances, filamentation is also likely to occur. The phase pre-correction therefore requires a precise knowledge of this self-induced spatio-temporal phase modulation in filaments.  

In the present work, we propose to make use of the unique properties of filamentation (self-broadening of the laser spectrum, self-cleaning of the spatial mode, temporal effects like self steepening,..) \cite{BraunKLDSM1995,CouaironM07,Chin2005,Kasparian} as part of the waveform synthesizer required to control molecules at a target distance, instead of trying to prevent it. Indeed, filamentation allows delivering  extremely high intensities remotely, circumventing the diffraction limit \cite{Stelmaszczyk_2004}. It is therefore well suited to initiate, control, and optimize non-linear processes, that require high-intensity, shaped pulses on the target.

Such approach requires to tailor a pulse so that it acquires the desired shape at the remote location of interest, i.e., during or after filamentation. 
In spite of intensity clamping~\cite{KaspaSC2000} and of the extreme phase sensitivity at the non-linear focus~\cite{Gaeta_phase}, we recently showed that filamentation is reversible \cite{Opex_retropropagation}. It is then possible to numerically back-propagate a target electric field to the initial conditions giving rise to it. Here, we show that this previously unexpected reversibility can be applied to explicitly design initial pulse shapes which yield the required target electric field for specific applications after filamentation or, more generally, after non-linear propagation. We illustrate this approach and its versatility by two examples featuring the shaping of temporal intensity, and spectral phase, respectively.

\section{Numerical Method}
\subsection{Non-linear propagation of ultrashort laser pulses}

The reversibility of laser filamentation relies on the inversion of the propagation direction in the governing equations\cite{Opex_retropropagation}, i.e., the non-linear Schr\"odinger equation (NLSE), or the unidirectional pulse propagation equation (UPPE)\cite{KolesMM2002}. These equations describe the propagation of the complex electric field $\varepsilon$ along the axis $z$ (assuming axisymmetry): 
\begin{equation}
\partial_{z}\widetilde{\varepsilon}= \mathrm{i}(k_{z}-\dfrac{\omega}{v_\mathrm{g}}) \widetilde{\varepsilon} + \dfrac{\omega}{c^{2}k_{z}} \bigg[\textrm{i}\omega 
n_{2} \widetilde{|\varepsilon|^{2}\varepsilon} 
-\dfrac{e^{2}}{2\epsilon_{0} m_\mathrm{e}} \tau(\omega)\ \widetilde{\rho \ \varepsilon} \bigg] - \widetilde{L[\varepsilon]},
\label{eq_UPPE}
\end{equation}
where $v_\mathrm{g}$ is the group velocity, $e$ and $m_{\textrm{e}}$ the charge and mass of the electron, respectively, $k_{z}=\sqrt{\textrm{k}^{2}(\omega)-\textrm{k}_{\bot}^{2}}$, with $k(\omega)$ being the wave vector and $k_{\bot}$ its transverse component, and $n_{2}$ the Kerr index. $\tau(\omega) = (\nu_{\mathrm{en}}+ \mathrm{i}\omega) / (\nu_{\mathrm{en}}^{2}+\omega^{2})$, $\nu_{\mathrm{en}}$ is the collision frequency between free electrons and neutral atoms and the tilde denotes a Fourier-Hankel transform:
\begin{equation}
\widetilde \varepsilon(k_{\bot},z,\omega)=\iint r \mathrm{J}_{0}(k_{\bot}r) \varepsilon(r,z,t)\mathrm{e}^{\mathrm{i}\omega t} \mathrm{d}t \mathrm{d}r,
\end{equation}
where $\mathrm{J}_{0}$ is the zeroth order Bessel function. The free-electron density $\rho$ evolves as:
\begin{equation}
	\partial_{t} \rho = \textrm{W}(|\varepsilon|^{2})(N - \rho) + \dfrac{\sigma}{U_{i}}|\varepsilon|^{2} - g(\rho),
\end{equation}
where $\textrm{W}(|\varepsilon|^{2})$ describes the probability of ionization calculated with the Perelomov, Popov, Terentev (PPT) formula \cite{PPT}, $\sigma$ is the inverse bremsstrahlung cross-section, $U_{\textrm{i}}$ the ionization potential, $N$ is the number density of  molecules, and $g$ is the recombination function. The last term in equation~(\ref{eq_UPPE}) accounts for ionization-induced losses, and is calculated as
\begin{equation}
L[\varepsilon]=\frac{U_{\mathrm{i}}W(|\varepsilon|^{2})}{2|\varepsilon|^{2}}(N-\rho)\varepsilon.
\end{equation}
At 400~nm, plasma constants are given in ref. \cite{CouaironM07} and the Kerr index in ref. \cite{Steinmeyer_n2}. At 800~nm, these constants are given in ref. \cite{Opex_retropropagation}.

\subsection{Back-propagation of filamenting pulses}
The pulses bearing the desired shapes (as  determined from prior theoretical arguments or laboratory experiments) at the target location are back-propagated by changing $\mathrm{d}z$ to $-\mathrm{d}z$ in the propagation equation (\ref{eq_UPPE}), as detailed earlier \cite{Opex_retropropagation}.

Reversibility however neither ensures that any arbitrary pulse shape can readily be back-propagated, nor that it will follow a filamentary path during its back-propagation. In other words, designing an initial pulse shape that will synthesize a target waveform after undergoing filamentation requires specific features.
For instance, to back-propagating arbitrary pulses can lead to intensity divergence due to the gain induced by the inversion of the loss term related to ionization \cite{Opex_retropropagation}.

\begin{figure}[t!]
	\centerline{
		\includegraphics[width=13cm]{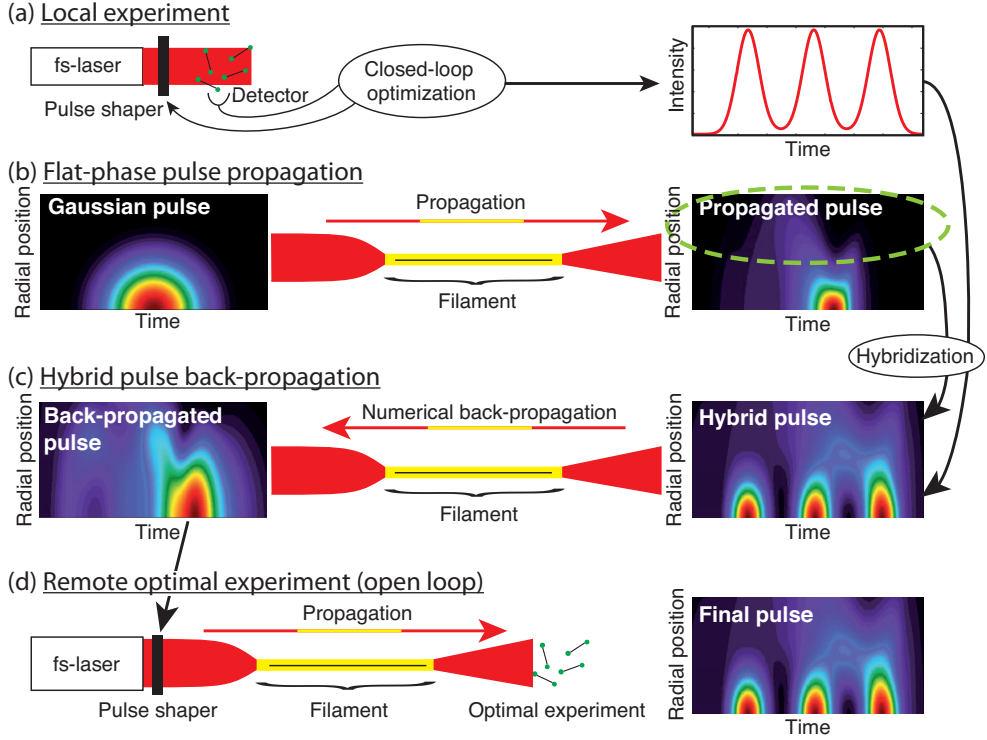}
			   }	
	\caption{(Color online) Principle of pulse hybridization for designing remote, post-filamentation pulse shapes for standoff detection applications. (a) Local \textit{a priori} calibration experiment defining the target pulse shape; (b) Forward filamentation propagation defining a pulse framework able to yield filamentation; (c) Back-propagation of the hybridized pulse, that mixes the two previous ones and defines the required initial pulse shape; and (d) Remote optimal experiment. The forward propagation of the hybridized pulse yields a shape identical to the target on the beam axis, thanks to the numerical stability of the filamentation reversibility.  \label{principe}}

\end{figure}

\subsection{Pulse hybridization}
Filamentation is likely to occur for high intensities and long propagation distances so that strong coupling between the spatial and temporal phase takes place. For this reason, we had to specifically design hybrid pulses encompassing the drastic spatial deformation induced by filamentation.
To design such pulses, we developed a hybridization strategy. As sketched in Figure~\ref{principe}, this is achieved by embedding the desired target shape (Panel~a) into the center of a post-filamentation propagated pulse (Panel~b), therefore resulting in backward filamentation (Panel~c) while bearing the tailored shape.
The size of the transition region has to be defined as a trade-off between three constraints: (i) ensure that the central region of the pulse is adequately shaped, (ii) provide a filamenting reference over a peripherial region bearing a substantial part of the pulse energy, and (iii) offer a sufficient intermediate zone to ensure a smooth transition.

More specifically, hybridization was performed in two distinct ways for temporal intensity and spectral phase shaping, respectively. 
In the former case, we applied  the target electric field $\varepsilon_{\textrm{t}}$ (Figure~\ref{principe}a) at the center of the beam profile, and progressively mixed it over the central region
 ($r \le r_\textrm{m}$~=~0.75~mm) with a reference field ($\varepsilon_{\textrm{p}}$ of phase $\Phi_\textrm{p}$) originating from the forward filament propagation (Figure~\ref{principe}b): 
\begin{equation}
	\varepsilon (r,t) = \\ \Big[ \varepsilon_{\textrm{p}}(r,t) \sin(\frac{r\pi}{2r_{\textrm{m}}} ) + \varepsilon_{\textrm{t}}(t) \Big(1-\sin( \frac{r\pi}{2r_{\textrm{m}}} )\Big)\Big] \textrm{e}^{\textrm{i}\Phi_{\textrm{p}}(r,t)}
\end{equation}
where the sine weighting was chosen so as to ensure a continuously differentiable radial transition between the two pulse components. 
Beyond $r_\textrm{m}$, 
the propagated pulse amplitude  $\varepsilon_{\textrm{p}}$ and phase $\Phi_{\textrm{p}}$ are used to maintain the broad spectrum generated by the forward-propagated filament. Such hybridization ensures the stable back-propagation as a filament. 
Simultaneously, the post-filamentation shape is preserved in the outer region of the profile. It accounts for 82\% of the pulse energy, ensuring the occurrence of filamentation in the backward propagation.

In the case of spectral phase shaping, we aimed at preserving simultaneoulsy the radial continuity of the electric field envelope in the temporal domain, and the 
wavefront curvature at each wavelength and for each radial position. We therefore hybridized the phase rather than the complex electric field. More precisely, we considered a propagated (post-filamentation) pulse of phase $\phi_\textrm{p}(r,\omega)$.
At each wavelength, the difference between the propagated phase on the side ($\phi_\textrm{p}(r,\omega)$) and the center ($\phi_\textrm{p}(0,\omega)$) describes the  radial phase dependence, and the associated wavefront curvature. 
The desired phase $\phi_\textrm{t}(\omega)$ was added to this curvature, defining in the Fourier space the target pulse to be backpropagated as:

\begin{equation}
	\widetilde{\varepsilon}(r,\omega) = \widetilde{\varepsilon}_{\textrm{p}} (r,\omega) \  \textrm{exp} \Big[ \textrm{i}  \Big(\phi_\textrm{t}(\omega) + \phi_\textrm{p}(r,\omega)- \phi_\textrm{p}(0,\omega) \Big)\Big].
\end{equation}

As hybridization affects the peak power and energy of the pulse, the latter was adjusted so as to recover a power allowing backward filamentation without triggering intensity divergence by the backward gain. 

The need to hybridize the electromagnetic waveform is illustrated in Figure \ref{triple_pulse_timeEvo}a. An 800~nm, 0.44 mJ non-hybridized shaped pulse, with a purely spherical wavefront curvature (blue dash-dotted line) does not enter a filamentation regime when back-propagated. In spite of a peak intensity of 70~TW$\,$cm$^{-2}$ in the focal region, self-guiding does not occur, as evidenced by the short length of both the high-intensity region (5~cm) and the ionized region (6~cm above 10$^{14}$~cm$^{-3}$, see inset), similar to the double Rayleigh length (5~cm) associated with a Gaussian pulse with identical beam diameter and focusing. A similar, filament-free backward-propagation dynamics is also obtained with longer focal lengths ($f$~=~2 and 3~m). This behavior contrasts in the case of back-propagation of a properly hybridized pulse (black line), the intensity of which is clamped at the same intensity of 70 TW$\,$cm$^{-2}$, showing that it is guided over 20~cm, and ionized over 17~cm.

\section{Results}

\subsection{Remote synthesis of pulses with temporal intensity control}

We first investigated the design of a pulse that will provide, after propagation, a desired temporal intensity shaping. We simulated the remote generation of a train of three Gaussian sub-pulses of 12~fs duration (FWHM)  at 800~nm, with 32~fs separation.This kind of structure is typical, for instance, for stimulating vibrational Raman scattering \cite{Kapteyn2003}.
This target pulse was back-propagated to $z = 0$ in order to define the initial pulseshape required to generate this target triple pulse after filamentation.

\begin{figure}[t!]
	\centerline{
		\includegraphics[width=13cm]
		{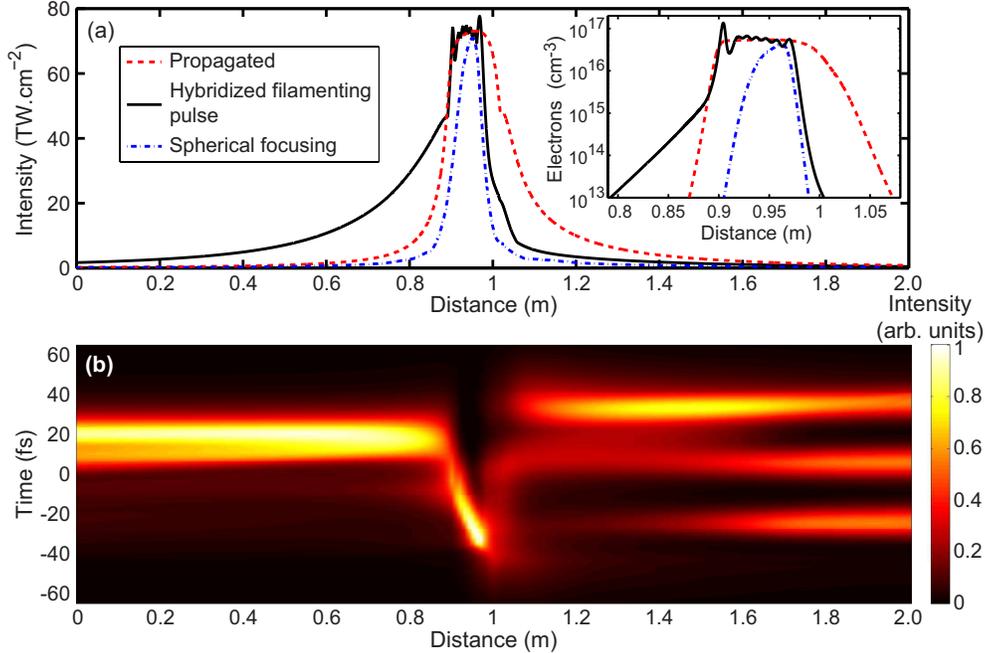}
			   }	
	\caption{(Color online) Back-propagation of a target triple pulse. (a) On-axis intensity and electron density ; (b) Evolution of the temporal dynamics of the pulse along the propagation distance. \label{triple_pulse_timeEvo}}
\end{figure}

\begin{figure}[t!]
	\centerline{
		\includegraphics[width=13cm]{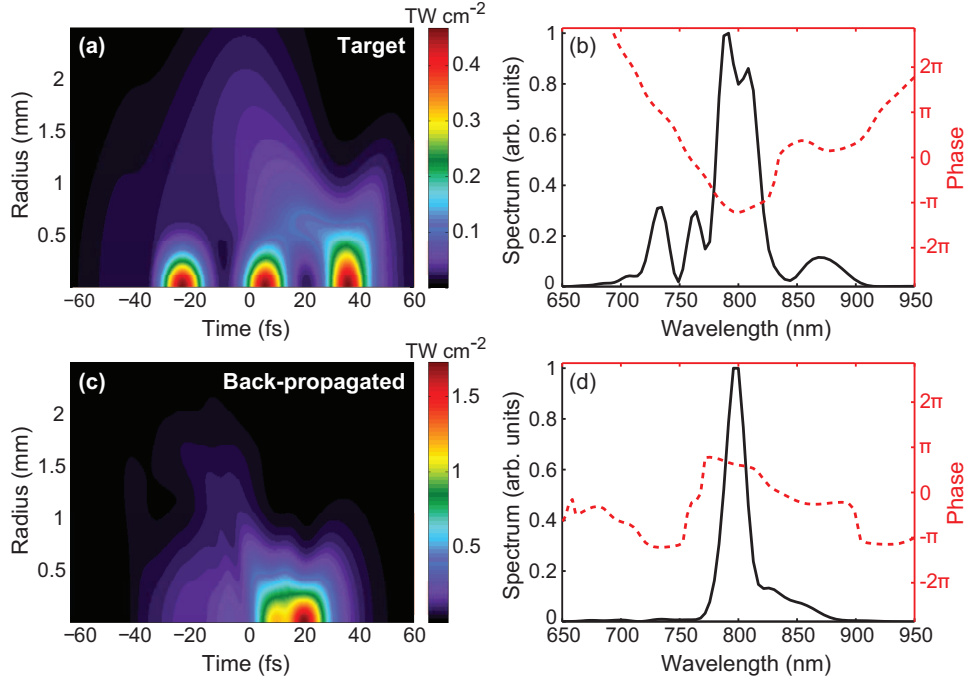}		
			   }	
	\caption{(Color online) Tailoring a triple pulse at 800~nm after laser filamentation. (a) Hybridized target pulse, featuring a triple pulse around its center, surrounded by the output of a regular filamenting pulse; (b) Corresponding on-axis spectrum and spectral phase on the beam axis; (c,d) Same data for the back-propagated pulse, i.e., the pulse shape that has to be propagated to get the target shape. \label{triple_pulse_Irt_Phase}}
\end{figure}

The target and back-propagated input pulses are notably different (Fig.~\ref{triple_pulse_Irt_Phase}). The spectrum  shrinks during the back-propagation, narrowing down to the spectral support for a Fourier Transformed 30 fs pulse envelope, while the temporal shape simplifies into a single peak envelope with a shoulder. Accordingly, the spectral phase is sufficiently smooth to be synthesized experimentally by a standard pulse shaper \cite{Weiner2000}.
The manipulation can therefore be seen as a controlled pulse splitting to generate a train of 3 Gaussian subpulses at the target location. The error between the target pulse (Fig.~\ref{triple_pulse_Irt_Phase}a) and the pulse obtained by numerically forward-propagating again the back-propagated pulse (Fig.~\ref{triple_pulse_Irt_Phase}c) stays below 0.2 \%, suggesting that the proposed non-linear pulse manipulation could be implemented experimentally.

The significant reshaping of the pulse is related to the complex dynamics occurring during filamentation (Figure~\ref{triple_pulse_timeEvo}b): Between $z$~=~0.85 and 1.05 m, the intensity reaches up to 70~TW$\,$cm$^{-2}$, the pulse maximum shifts temporally, and then splits into three sub-pulses, two of them developing in the far-field during the post-filamentation propagation stage. This complex evolution of the pulse illustrates the rich dynamics of filamentation, and the associated difficulty of explicit pulse design. It also demonstrates the relevance of the proposed pulse hybridization procedure: It adequately prepares pulseshapes such that filamentation acts as a complementary nonlinear pulse shaping step.
From a more fundamental point of view, in this example, tailoring only 18\% of the pulse energy has a crucial influence on the final pulse shape, as expected due to the high sensitivity of non-linear propagation to the initial conditions. 

\subsection{Remote synthesis of pulses with spectral phase control}
The temporal hybridization used in the above example, although intuitive, leads to an inhomogeneous temporal profile along the radius of the back-propagated pulse (Figure \ref{triple_pulse_Irt_Phase}c), a shape that cannot be realized experimentally with today's technology. In contrast, the phase hybridization can be expected to result in a more homogeneous temporal shape along the radius.
Furthermore, phase shaping is the most widely used approach for coherent control, e.g. in view of optical discrimination of molecules featuring similar spectroscopic properties \cite{Assion_science1998}.



\begin{figure}[t!]
	\center{
		\includegraphics[width=13cm]{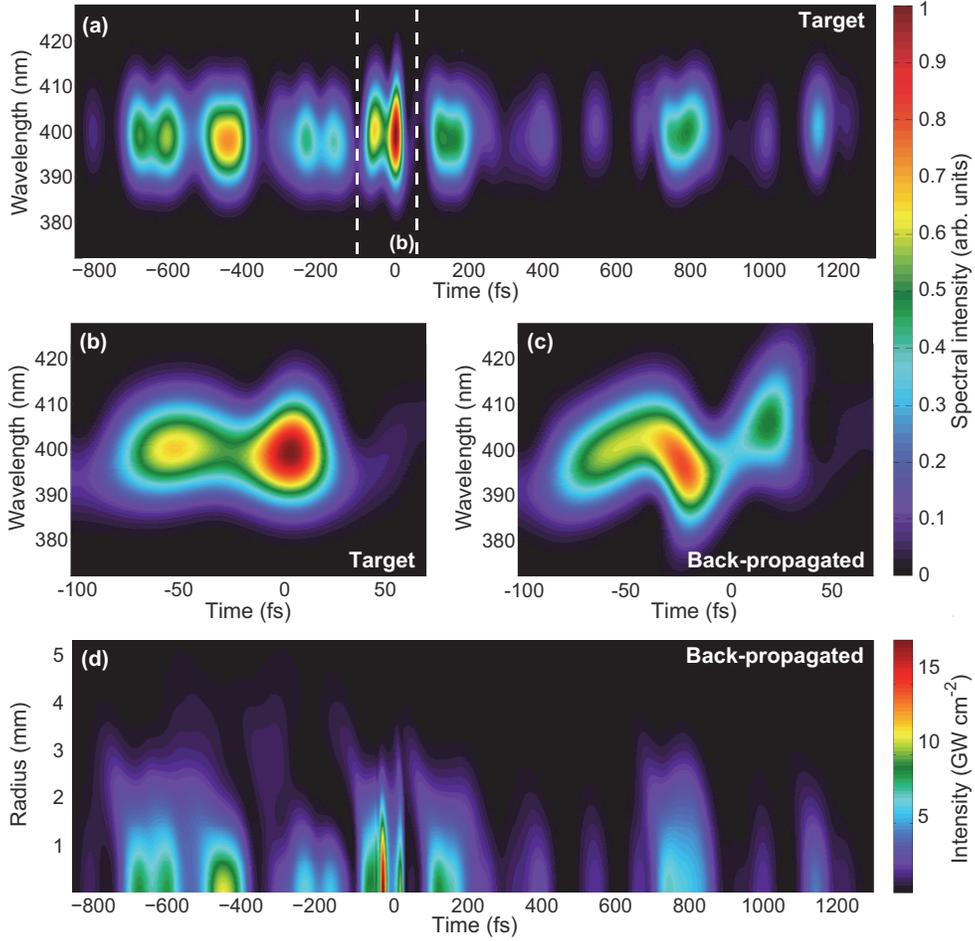}		
			}	
	\caption{(Color online) Tailoring of an optimally phase-shaped pulse for remote selective detection of RBF with regard to FMN:  Wigner plots (with 12~fs FWHM Gaussian gate) of (a) the target pulse shape, and the central part of (b) the target and (c) the back-propagated pulses; (d) Radial distribution of the intensity of the back-propagated pulse. \label{OptP_wigner}}
\end{figure}

With highly selective standoff detection applications in mind, we investigate the possibility of discriminating the presence of molecules relevant to bioaerosols: Riboflavin (RBF) and flavin mononucleotide (FMN) \cite{M'ejKYFSW2004}. This demonstration was successfully performed in the laboratory \cite{Boutou_FMN,Petersen2010}, in which the optimal waveforms were determined. Figure~\ref{OptP_wigner}a illustrates the complexity of this optimal pulse, featuring seven significant sub-pulses with different chirps and temporal shapes.
Realizing this biomolecular identification at a distance implies delivering these high-intensity pulses at 400~nm with the spectral phase maximizing the depletion of fluorescence from RBF as compared to that of FMN (and \textit{vice-versa}).

\begin{figure}[t!]
	\centerline{
		\includegraphics[width=13cm]{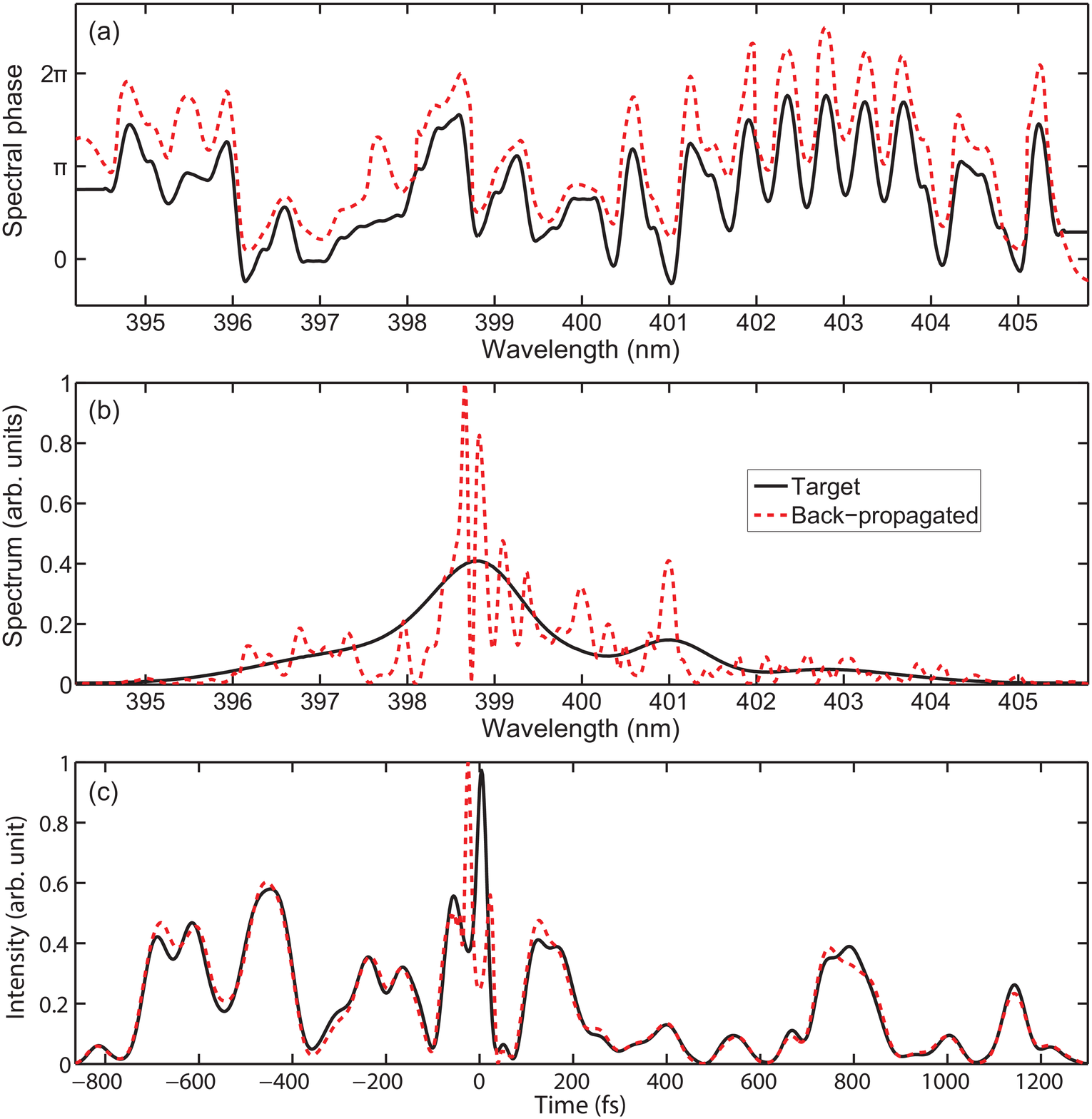}
			   }	
	\caption{(Color online) Target and back-propagated waveforms on-axis (a) spectral phase, (b) spectrum and (c) temporal pulse shapes of an optimally phase-shaped pulse for remote selective detection of FMN with regard to RBF. \label{OptP_phase}}
\end{figure}

The contrast between target and back-propagated pulses is clearly observed on both the spectrum and the spectral phase. The target and initial pulses exhibit differences up to 60 \% for the central peak intensity, and up to 0.9~$\pi$ for the dephasing of the spectral phase (Figure \ref{OptP_phase}a,b).  In the temporal domain (Figure \ref{OptP_phase}c), this results in a splitting of the central peak of the back-propagated pulse  (between $t$~=~-100 and 60 fs). These deformations are also visible in both the temporal and frequency spaces, as evidenced by the Wigner plot at the center of the pulse (Figure~\ref{OptP_wigner}b,c). While the target pulse has a double structure without chirp (Figure~\ref{OptP_wigner}b), the above-mentioned splitting results in three sub-pulses in the back-propagated pulse (Figure~\ref{OptP_wigner}c). Furthermore, the two first ones have opposite chirps.
In the temporal domain, the pulse reshaping mainly occurs on phase rather than on intensity, as evidenced by Figure~\ref{OptP_phase}c.

The shape of the back-propagated pulse depends much less on the radial position than with the temporal hybridization (Fig \ref{OptP_wigner}d). This homogeneity appears even more clearly when considering that the beam is more than twice as large as in the previous case, and that the phase curvature is very small on the first two millimetres around the axis.

\begin{figure}[t!]
	\centerline{
		\includegraphics[width=13cm]{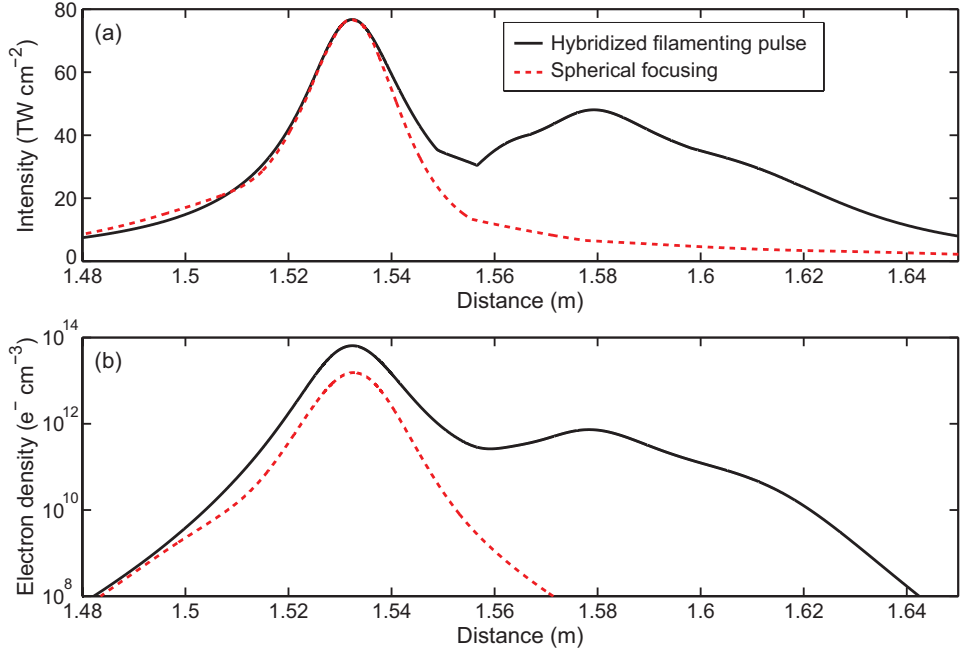}
			   }	
	\caption{(Color online) (a) Intensity and (b) free electron density dynamics during the back-propagation of a spherically focused pulse at 400~nm, and of a hybridized filamenting pulse optimizing the discrimination of RBF \textit{vs.} FMN. \label{OptP_IntDist}}
\end{figure}

The spectacular reshaping of the pulse during the back-propagation is again related to the phase scrambling induced by non-linear propagation associated with filamentation. As illustrated on Figure~\ref{OptP_IntDist}, the back-propagation of our target pulse displays filamentation over almost 10~cm, as evidenced by the ionized section and the refocusing cycle. 
This is again made possible by preparing an adequate pulse hybridization.
We compared these results with the back-propagation of a 0.84~mJ pulse  with the same spectral phase but a spherical wavefront curvature instead of pulse hybridization ensuring filamentation during the back-propagation. The non-hybridized pulse does not enter a filamentation regime when back-propagated, although it reaches the same peak intensity of 76~TW$\,$cm$^{-2}$ in the focal  region (1.6~m, Figure \ref{OptP_IntDist}).

\
\section{Discussion}
The results unambiguously show the possibility of explicitly tailoring the output of laser filaments in air to \textit{a priori} arbitrary waveforms. Potential applications range from selective standoff detection \cite{Wipfler2012,NatanLGKS2012,Bremer2011,Bremer2013,Hemmer2011,Ariunbold2014,Glenn2014,M'ejKYFSW2004,Gravel2004,Boutou_FMN2,ScullyPNAS_2011,ScullyPNAS_2012} to lightning control \cite{KaspaAAMMPRSSYMSWW2008a}, atmospheric water condensation \cite{Henin2011}, and radiative forcing modulation \cite{Leisner_PNAS2013}. Similar techniques could in principle be applied to any other remote application. It only requires that   the target shape can be explicitly designed and/or empirically determined \textit{a priori}, e.g., via a closed-loop algorithm or other optimization techniques in laboratory (local) experiments beforehand, or from numerical simulations of the phenomenon of interest. For example, we successfully back-propagated several other target waveforms, including Gaussian pulses, double pulses, or phase controlled pulses maximizing FMN \textit{vs.} RBF fluorescence depletion instead of minimizing it. 

The experimental implementation of the pulse tailoring technique proposed in the present work requires the back-propagated pulse shapes to be compatible with present technical constraints, especially a decoupling of the spatial and spectral phases. Our work shows that the way to perform the pulse hybridization is critical in that regard.

Finally, the proposed pulse design technique is based on the reversibility of filamentation, that does not depend on the specific filamentation or ionization model \cite{Opex_retropropagation}. Consequently, our technique is also independent on these models.

\section{Conclusion}
In summary, we have demonstrated the explicit remote tailoring of arbitrary pulses after non-linear propagation, and more specifically at the end of laser filamentation. It opens the way to applications requiring the remote delivery of specific laser waveforms with high intensity, in order to optimize atmospheric applications like lightning control or laser-induced condensation, or selective remote sensing. 

\begin{acknowledgments}
We acknowledge financial support by the European Research Council Advanced Grant ``Filatmo'', the Swiss national foundation NCCR MUST program, and the Marie Curie COFUND program of FP7.
Discussions with Pierre B\'ejot (Universit\'e de Bourgogne) were very rewarding and the assistance of Michel Moret was highly appreciated. 
\end{acknowledgments}

\end{document}